\documentstyle{EuroPhys}
\input EuroMacr
\input epsf
\newcommand{\hmp}{h^{-1} Mpc }
\newcommand{\be}{\begin{equation}}
\newcommand{\ee}{\end{equation}}
\def\spose#1{\hbox to 0pt{#1\hss}}
\def\ltapprox{\mathrel{\spose{\lower
3pt\hbox{$\mathchar"218$}}
 \raise 2.0pt\hbox{$\mathchar"13C$}}}
\def\gtapprox{\mathrel{\spose{\lower
3pt\hbox{$\mathchar"218$}}
 \raise 2.0pt\hbox{$\mathchar"13E$}}}
\def\inapprox{\mathrel{\spose{\lower
3pt\hbox{$\mathchar"218$}}
 \raise 2.0pt\hbox{$\mathchar"232$}}}

\begin{document}
\shorttitle{M.MONTUORI \etal Galaxy number counts and Fractal correlations}

\title{Galaxy number counts and Fractal correlations }
\author{M. Montuori \inst{1},F. Sylos Labini\inst{1},
   A. Gabrielli\inst{1},A. Amici\inst{1} \And L. Pietronero\inst{1}}
\institute{
     \inst{1} INFM Sezione di Roma1, and
                Dipartimento di Fisica, Universit\`a di 
           ``La Sapienza''Department of   Physics -
           P.le A. Moro 2, I-00185 Roma, Italy\\
    }
\rec{ }{ }
\pacs{
\Pacs{98}{65$-$r}{large scale structure of the Universe}
\Pacs{05}{20$-$y}{Statistical Mechanics}
      }
\maketitle

\begin{abstract}
We report the correlation analysis of various redshift surveys 
which shows that the available data are consistent with each other 
and manifest fractal correlations (with dimension $D \simeq 2$) 
up to the present observational limits ($\approx 150 \hmp$) 
without any tendency towards homogenization.
This result points to a new interpretation of 
the number counts that represents the main subject 
of this letter.
We show that an analysis of the small scale fluctuations allows us 
to reconcile the correlation analysis and the number counts in a 
new perspective which has a number of important implications.
\end{abstract} 


Ideally the study of the correlation analysis of 
galaxy distribution requires the knowledge of the 
position of all galaxies in space
 \cite{dp83} \cite{cp92}.
In practice, the observation of angular positions plus the redshift
provides a redshift catalogue in which galaxies are located in the
 three
dimensional  space, but
such a catalogue is affected by a luminosity selection
effect related to the observational point. In order to avoid this
 effect, one
can define a maximum depth and include in the sample only those
 galaxies that
would be visible from any point of this volume. This procedure 
defines a volume limited 
(VL) sample, whose statistical  properties  are  
unaffected by observational biases \cite{dp83} \cite{cp92}.

We discuss here the determination of the space density in 
various redshift and angular surveys. The underlying assumption used 
is that the space $\rho(r)$ and 
luminosity $\phi(L)$ distributions 
are independent \cite{bst88}. In such a way the number of galaxies for unit 
luminosity and unit volume can be written as 
$\nu(L,r) d^3r dL= \rho(r)d^3r \phi(L) dL$. Although this 
assumption  is not strictly valid in view of the correlation
between galaxy positions and (absolute) luminosities,
for the purpose of the present discussion this approximation 
is rather good \cite{slp96}.

We start recalling  the 
concept of correlation. If the presence of an object at the point $r_1$ 
influences the probability of finding another object 
at $r_2$, 
these two points are correlated. Therefore there is a correlation
at  $r$ if, on average
$G(r) = \langle n(0)n(r)\rangle   \ne \langle n\rangle ^2 
$,
where we average over all occupied points chosen as origin.
On the other hand, there is no correlation at $r$ if
$G(r) \approx \langle n\rangle  ^2$.
The length scale $\lambda_0$, 
which separates correlated regimes from
uncorrelated ones, is the homogeneity scale.

In the analysis, it is useful to 
use \cite{cp92}
$\Gamma(r)  
= G(r)/<n>$
where $\:<n>$ is the average density of the sample analyzed.  
The reason is that $\Gamma(r)$ has an amplitude independent 
from the sample size, differently from $G(r)$, and it is 
suitable for the comparison between different samples.

$\Gamma(r)$ 
can be computed by the following expression:
\be
\label{e4}
\Gamma(r) = \frac{1}{N} \sum_{i=1}^{N} \frac{1}{4 \pi r^2 \Delta r}
\int_{r}^{r+\Delta r} n(\vec{r}_i+\vec{r'})d\vec{r'} = 
\frac{BD}{4 \pi} r^{D-3}
\ee
where $D$ is the fractal dimension and $B$ is the lower cut-off (see below).
$\Gamma(r)$ is the average density at distance $\:\vec{r}$ from an
occupied point at $\vec{r_i}$
and it is called the {\it conditional average density} \cite{cp92}.
If the distribution is fractal up to a certain distance $\lambda_0$,
and then it becomes homogeneous, $\Gamma(r)$ 
is a power law function of $r$ up to $\lambda_0$, and then it flattens 
to a constant value.
Hence by studying the behavior of $\Gamma(r)$
it is possible to detect the eventual scale-invariant properties
of the sample. Instead the information given by the standard correlation
function  $\xi(r)$ 
\cite{dp83}  \cite{pe93}
 is biased by the 
a priori (untested) assumption of homogeneity \cite{cp92}.

Given a certain sample with solid angle $\Omega$ and depth $R_{s}$,
it is important to define which is 
 the maximum distance up to which it 
is statistically meaningful 
to compute the correlation function. 
As discussed in \cite{cp92}, 
the conditional density $\Gamma(r)$ 
has to be computed in spherical shells; in this way we do not make 
any assumption in the treatment of
the boundaries conditions.
For this reason, the maximum distance up 
to which we extend our analysis is the order 
of the radius $R_{eff}$ of the largest 
sphere fully contained in the sample volume.
In such a way  we do not consider in the statistics
the points for which a sphere of radius {\it r} is not
fully included within the sample boundaries.
For this reason we have a smaller number of points
and we stop our analysis  at a  shorter depth than 
other authors ones.

When one evaluates the correlation
function
(or the  power spectrum \cite{sla96}) beyond $R_{eff}$,
then one  makes explicit assumptions on what
lies beyond the sample's boundary. In fact, even in absence of
corrections for selection effects, one
is forced to consider incomplete shells
calculating $\Gamma(r)$ for $r>R_{eff}$,
thereby
implicitly assuming that what one  does not find  in the part of the
shell not included in the sample is equal to what is inside.

We show in Fig.1 
the determination of the conditional  density in VL with the same
cut in absolute magnitude, in 
different surveys (see \cite{slmp97} 
for a review on the subject).
The match of the amplitudes and exponents
is quite good. 
The main result is that galaxy distribution shows fractal correlations with
 dimension
$D \approx 2$ up to the limiting length $R_{eff}$, which is 
different for the various samples (ranging from $20 \hmp$ to about $150 \hmp$) 
\cite{cp92} \cite{slmp97}.
   There have been attempts to push $R_{eff}$ to larger values by using
 various
weighting schemes for the treatment of boundary conditions
 \cite{gu92}.
 These methods
however, unavoidably introduce artificial homogenization effects and 
therefore
should be avoided \cite{cp92}. 
A different way to get information for larger scales is 
presented in the following.

  Historically \cite{pe93},
the oldest type of data about galaxy distribution is given by
the relation between the 
number of observed galaxies $N(>f)$  and their apparent
brightness $f$. It is easy to show that \cite{pe93}
$N( >f)  \sim  f^{-\frac{D}{2}}$
where $D$
 is the fractal dimension of the galaxy distribution. 
 In terms of the apparent magnitude 
$f \sim 10^{-0.4 m}$ 
(note that bright galaxies correspond to small $m$), 
the previous relation becomes
$\log N(<m)   \sim \alpha m$ with $\alpha = D/5$ 
 \cite{pe93}. In Fig.2 we have collected all the
recent observations of $N(<m)$ 
versus $m$ \cite{slgmp96}.
 One can see that at small scales (small
$m$) the exponent is $\alpha \approx 0.6$, while at larger scales 
(large $m$) it
changes into $\alpha  \approx 0.4$. The usual interpretation
\cite{pe93}   is that $\alpha \approx  0.6$
corresponds to $D \approx 3$ consistent with homogeneity, while 
$\alpha \approx  0.4$
 is the result of  large scales
galaxy evolution and space time expansion effects.
On the basis of the previous discussion of the VL samples, we can 
see that this
interpretation is untenable. In 
fact, there are very clear evidences that, at least up 
to $150  \hmp $
there are fractal correlations \cite{cp92} \cite{slmp96},
 so one would eventually expect the
opposite behavior: 
$\alpha \approx 0.4$ (fractal with  $D \approx 2$) for
small $m$,  and $\alpha  \approx 0.6$ for large  $m$. 
  An additional argument addressed in favor of homogeneity, 
at rather small 
scales, is the rescaling of angular 
correlations \cite{pe93}.
This again seems to be in contradiction with the properties 
observed in 
the VL
correlation analysis.

We  show that this contradictory situation arises from the 
fact that,
given the limited amount of statistical information corresponding to 
the
various methods of analysis, only some of them can be considered 
as
statistically valid, while others are strongly affected by finite 
size and other 
spurious fluctuations that may be confused with real 
homogenization \cite{slgmp96}.
 We focus now on the possibility of
extending the sample effective depth $R_{eff}$.
 In order to
discuss this question,
 it is important to analyze the properties
 of the {\it  small scale
fluctuations}.  
To this aim, we introduce the conditional density in 
the 
volume $V(r)$ as {\it observed from the
 origin}, defined as 
 \be 
\label{new1} 
n(r) = \frac{N(<r)}{V(r)} = \frac{3Bp}{4\pi} r^{D-3}  \; .
\ee  
In principle  Eq.\ref{new1} should refer  to {\it all}
the galaxies present  in the 
volume $V(R)$. If instead we have a VL sample,
 we will see only a fraction
$N_{VL}(R) = p \cdot N(<R)$  (where $p<1$)
 of the total number $N(<R)$ of galaxies in $V(R)$. 
If $\phi(L)dL$ is
 the fraction of galaxies whose absolute luminosity ($L$) is between 
$L$ and
$L+dL$ \cite{sc76}, $p$ is given: 
\begin{equation}
\label{p1}
 0< p = \frac{\int_{L_{VL}}^\infty  \phi(L)dL} 
{\int^\infty_{L_{min}} \phi(L)dL} < 1
\end{equation}
The function $\phi(L)$
 has been extensively measured \cite{dac94} 
and it is a
power law extending from a minimal value $L_{min}$ 
to a maximum value $L^*$
defined by an exponential cut-off. In Eq.\ref{p1}
$L_{VL}$ is the minimal absolute luminosity that 
characterizes the VL
sample and $L_{min}$ is 
the fainter absolute luminosity (or magnitude $M_{min}$)
surveyed in the catalog (usually $M_{min} \sim -11 $).
 Computing $n(r)$, we expect (Fig.3 - insert panel) 
not to see any galaxy up to a certain 
distance $\ell_v$. 
For a Poisson distribution this distance is 
of order of the mean average distance between neighboring 
galaxies, $\ell_v \sim (V/N)^{\frac{1}{3}}$. Of course, 
such a quantity is not intrinsic for a
 fractal distribution because it 
depends  on the 
sample volume,  while the meaningful 
measure is the average minimum distance 
between neighboring galaxies $\ell_{min}$, that is related 
to the lower cut-off of the distribution. 
For distances somewhat 
larger than
$\ell_{min}$ we expect therefore a raise of 
the conditional density 
because we
are beginning to count some galaxies and 
$n(r)$ is affected by the fluctuations due to the 
low statistics. 
It is therefore important to be able to
estimate and control the {\it minimal
statistical length} $\lambda$, which separates 
the fluctuations due to the low statistics 
from the genuine behavior  of the distribution. 
A simple argument for the 
determination on the length $\lambda$
is the folliwng (see also \cite{slgmp96}).
At small scale, where there is a small number of galaxies, 
there is an additional 
term, due to shot noise, superimposed to the power
law behavior of $n(r)$, that destroys the genuine correlations 
of the system.  Such a fluctuating term can be 
erased out by making an average over all the points 
in the survey. 
On the contrary, in the observation from the origin, 
only when the 
number of galaxies is larger than, 
say, $\sim 30$, then the shot noise 
term can be not important.
 This condition gives (from Eq.\ref{new1})
\begin{equation}
\label{v3}
\lambda =  5 \left( \frac{4 \pi}{B p \Omega} \right)^{\frac{1}{D}} \approx 
\frac{20 \div 60 h^{-1} Mpc}{\Omega^{\frac{1}{D}}}. 
\end{equation}
for a typical VL sample with $M_{VL} \approx M^*$, where 
 $B$
 corresponds to the amplitude of the conditional density of all
galaxies \cite{slgmp96} \cite{slmp97}. 
This can be estimated from the amplitude of $\Gamma(r)$ 
in a VL sample 
divided by the correspondent $p$ as
defined in Eq.\ref{p1}.  We find  (for typical catalogues) 
$B \approx 10\div 15 (h^{-1} Mpc)^{-D}$ \cite{slgmp96}.

In Fig.3 we report the radial density 
estimated from the 
origin for
different VL samples derived from the PP catalogue. The finite size 
transient
behavior is evident and the correct scaling is reached for lengths 
larger than
$\lambda  \approx 50 \hmp$ ($\Omega = 0.9 sr$),
 the same for all the VL samples. In Fig.2 we
can see that this behavior is in perfect agreement with the full 
correlation
analysis corresponding to smaller scales.
In Table I we report the values of $\lambda$ for the various 
catalogues. We
have checked the validity of these values for the available
 catalogues (CfA1, PP,   SSRS1,
LEDA, ESP), as well as for artificial simulations as a test. Indeed in all
these catalogues one observes a well defined power law
for $R > \lambda$, 
corresponding to a
fractal dimension $D\approx 2$, up to the catalogue depth
\cite{slgmp96}. It is remarkable to note
that for the ESP catalogue this depth is  $\approx 
 800 \div 900  \hmp$ \cite{slmp97}.

The introduction of the {\it minimal statistical length}
 $\lambda$ has a very important
effect on the number counts $N(<m)$ and on the analysis of angular 
samples. For 
the number counts it is clear that if the majority of the galaxies in
 the survey are 
located at distances smaller than $\lambda$ this will not give us 
reliable
statistical information. In particular, the region up to $\lambda$ is
characterized by a strongly fluctuating regime,  followed by a decay 
just after
$\lambda$ (Fig.3 insert panel). For integral quantities as the number counts, 
such a
behavior can be roughly approximated by a constant conditional 
density over some range of scales. This
will lead to an apparent exponent $\alpha \approx
 0.6$ as if the distribution would be
really homogeneous. If instead the majority of galaxies lie in the
 region
beyond $\lambda$ the number counts will correspond to 
the real
statistical properties. 

 To be more quantitative, suppose to have a certain
 survey characterized by a solid angle $\Omega$  and we ask the
 following question: up to which apparent magnitude limit 
 $m_{lim}$ do we have to push  our observations to obtain that the
 majority of the galaxies lie in the statistically significant 
 region ($r \gtapprox  \lambda$) ?  Beyond
 this value of $m_{lim}$ we should
 recover the genuine properties
 of the sample because, as we have
 enough statistics, the finite
 size effects self-average out. From the previous condition for
 each solid angle $\Omega$ we can find an apparent magnitude limit
 $m_{lim}$.
  
To this aim, we can 
require that, in a ML sample, the peak of 
the selection function, which occurs at distance $r_{peak}$,
satisfies the condition
$r_{peak} > \lambda$
.
The peak of the survey 
selection function occurs for $M^* \approx -19$ 
 and then we have 
$r_{peak} \approx 10^{\frac{m_{lim}-6}{5}}$.
From the previous relation and Eq.\ref{v3}
we have that
\be
\label{cayy2}
m_{lim}= M^*-5\log(\lambda) + 25 \approx 14 -\frac{5}{D} \log(\Omega) \; .
\ee
It follows that for $m > 19$
 the statistically significant region  is reached for almost {\em
 any} reasonable value of the survey solid angle. This implies that
 in deep surveys, if we have enough statistics, we   readily 
 find the right behavior ($\alpha =D/5$), while it does not happens
 in a self-averaging way for the nearby samples. Hence the exponent
 $\alpha \approx 0.4$ found in the deep surveys ($m>19$) is a {\em
 genuine feature of  galaxy distribution}, and corresponds to real
 correlation properties.  
In the nearby surveys $m < 17$ we do not
 find the scaling region in the ML sample for almost {\em any}
reasonable
 value of the solid angle. Correspondingly the value of the
 exponent is subject to the finite size effects, and to recover the
 real statistical properties of the distribution one has to perform
 an average.  

We can now go back to Fig.2 and give to 
it a completely new 
interpretation. At
relatively small scales we observe 
$\alpha \approx 0.6$ just because of finite size
effects and not because of real homogeneity. This resolves the apparent
contradiction between the number counts and the 
correlation in  VL samples that show fractal
behavior up to $\sim 200 \hmp$. For
$m>19$ we are instead sampling a
distribution in which the majority of galaxies are at distances larger 
than
$\lambda$ and indeed $\alpha  \approx 0.4$, 
corresponding to $D \approx 2$, in full agreement
with the correlation analysis. Note that the change of slope 
at $m \approx  19$ depends
only weakly on the solid angle of the survey. In order to check that 
the
exponent $\alpha  \approx 
0.4$ is the real one we have made various tests on PP where
also one observes $\alpha  \approx 
0.6$ at small values of $m$, but we know that the
sample has fractal correlations from the complete space analysis
\cite{slgmp96}. An average of
the number counts from all points leads instead to the correct exponent
$\alpha \approx  0.4$ because for average quantities the 
effective value of $\lambda$ becomes actually appreciably
smaller (see
\cite{slgmp96} for more details). Our conclusion is therefore that 
there is not {\it any change of slope} at $m \sim 19$, and we 
see the same exponent in the range $ 12 \ltapprox m \ltapprox 18$,
where the combined effects K-corrections, galaxy evolution and
modification of the Euclidean geometry are certainly
negligible, and in the range 
 $ 19 \ltapprox m \ltapprox 28$. 

\section{Figures and Tables}

\begin{table}
\caption{In this table we summarize the characteristic
properties of several redshift catalogues and their volume limited
samples. $\Omega$ is the solid angle, $R_{VL}$ the depth of the VL sample 
and $N_{VL}$ the total number of galaxies. The {\it minimal statistical length}
 $\lambda$ 
gives us the scale above which the analysis of the conditional 
density from the origin is statistically meaningful.}
\[
\begin{tabular}{|c|c|c|c|c|}
\hline
	&	&	&	&	\\
\multicolumn{1}{c}{\rm{Survey}} &\multicolumn{1}{c}{$\Omega (sr)$} 
& \multicolumn{1}{c}{$\lambda (\hmp)$}  & \multicolumn{1}{c}{$R_{VL} (\hmp)$} & 
$N_{VL}$  \\
\hline
\multicolumn{1}{c}{CfA1} & \multicolumn{1}{c}{1.8} & 
\multicolumn{1}{c}{15}	          & \multicolumn{1}{c}{40}  & 442\\
\multicolumn{1}{c}{CfA2 (North)} & 
\multicolumn{1}{c}{1.3} & 
\multicolumn{1}{c}{20 }& \multicolumn{1}{c}{101} 
& 1031 \\
\multicolumn{1}{c}{PP} & 
\multicolumn{1}{c}{0.9} & 
\multicolumn{1}{c}{50}  & 
\multicolumn{1}{c}{60} & 990\\
\multicolumn{1}{c}{SSRS1} & 
\multicolumn{1}{c}{1.75} & 
\multicolumn{1}{c}{15}  & 
\multicolumn{1}{c}{60} & 345\\
\multicolumn{1}{c}{LEDA(m=16) }& 
\multicolumn{1}{c}{2 $\pi$ }	& 
\multicolumn{1}{c}{10} & 
\multicolumn{1}{c}{80} & 4550 \\
\multicolumn{1}{c}{IRAS1.2Jy }& 
\multicolumn{1}{c}{4 $\pi$} 	& 
\multicolumn{1}{c}{10} & 
\multicolumn{1}{c}{60} & 876\\
\multicolumn{1}{c}{ESP} & 
\multicolumn{1}{c}{0.006} 	& 300  &    &\\
\hline
\end{tabular}
\]
\label{pert1}
\end{table}

\newpage
\begin{figure}[h]
\epsfxsize=6cm
\epsffile{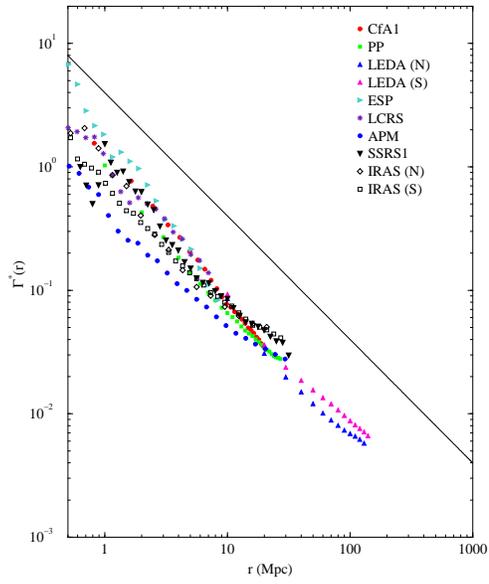}
\caption{ The spatial density   $\Gamma(r)$     computed in some 
VL samples of CfA1, PP, LEDA, APM, ESP, LCRS, SSRS1, IRAS 
and ESP
and normalized to the corresponding factor, as explained in the
text. 
}
\label{fig1}
\end{figure}

\begin{figure}[h]
\epsfxsize=6cm
\epsffile{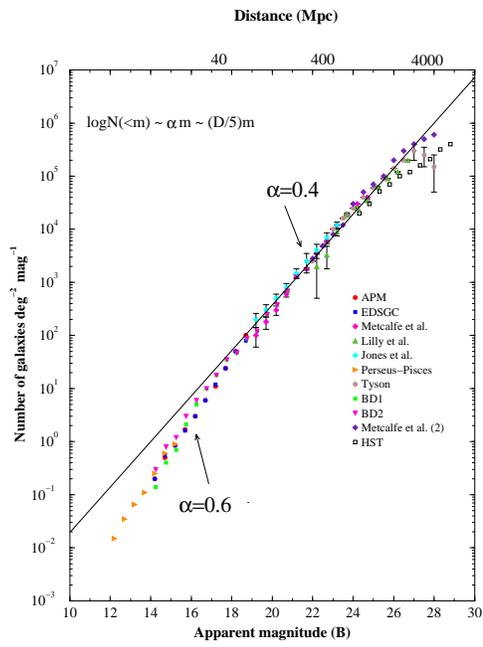}
\caption{Galaxy number counts as a 
function of the apparent magnitude ($m$) in the visible B-band. 
$\alpha = D/5 \approx 0.6$
}
\label{fig2}
\end{figure}

\begin{figure}[h]
\epsfxsize=6cm
\epsffile{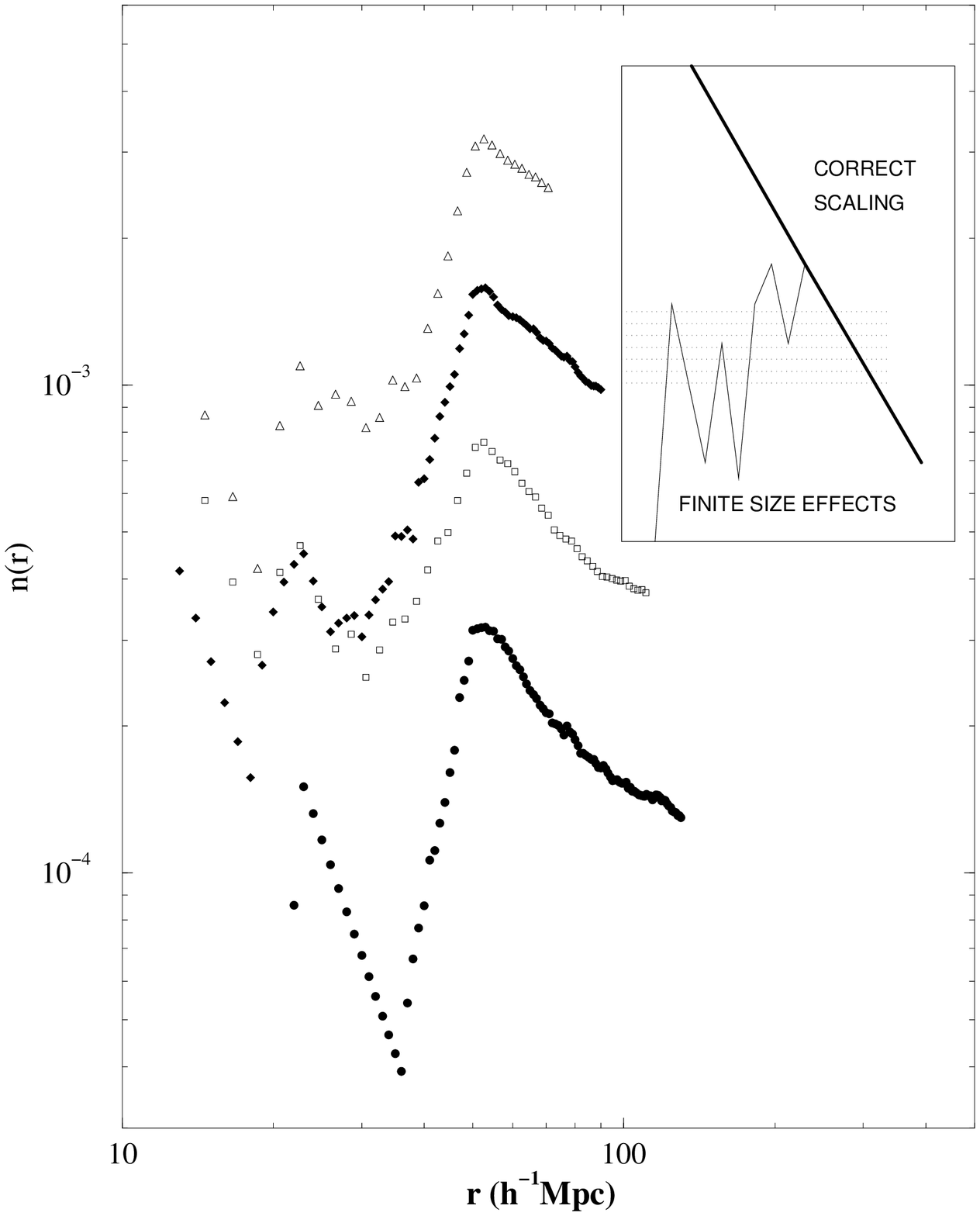}
\caption{Conditional density computed form one point in
various VL samples of PP. 
 The behavior of the average conditional density 
(Fig.2 Top panel) can be extended to larger scales by the 
conditional density form the vertex only for $r > \lambda$ 
where it becomes statistically meaningful. In the insert panel:
Schematic behavior of the conditional density
computed form a single point (the origin).
}
\label{fig3}
\end{figure}
\end{document}